\documentclass[epj]{webofc}
\usepackage[varg]{txfonts}   % Web of Conferences font
\usepackage{lineno}
\begin{document}
%
%\linenumbers
%
\title{Interpretation of astrophysical neutrinos observed by IceCube experiment by setting Galactic and extra-Galactic spectral components}
%
% subtitle is optionnal
%
%%%\subtitle{Do you have a subtitle?\\ If so, write it here}

\author{     
         Daniele Gaggero\inst{1} \and
         Dario Grasso\inst{2} \and
         Antonio Marinelli\inst{2}\fnsep\thanks{\email{antonio.marinelli@pi.infn.it}} \and
         Alfredo Urbano\inst{3} \and
         Mauro Valli\inst{4}
         }

\institute{GRAPPA Institute, University of Amsterdam, Science Park 904, 1090 GL Amsterdam, The Netherlands
\and
           INFN and Dipartimento di Fisica ``E. Fermi", Pisa University, Largo B. Pontecorvo 3, I-56127 Pisa, Italy
\and
           CERN, Theory division, CH-1211 Gen\`eve 23, Switzerland          
\and
           SISSA and INFN, via Bonomea 265, I-34136 Trieste, Italy           
          }

\abstract{%
The last IceCube catalog of High Energy Starting Events (HESE) obtained with a livetime of 1347 days comprises 54 neutrino events equally-distributed between the three families with energies between 25 TeV and few PeVs. Considering the 
homogeneous flavors distribution (1:1:1) and the spectral features of these neutrinos the IceCube collaboration claims the astrophysical origin of these events with more than $5\sigma$. The spatial distribution of cited events does not show a clear 
correlation with known astrophysical accelerators leaving opened both the Galactic and the extra-Galactic origin interpretations. Here, we compute the neutrino diffuse emission of our Galaxy on the basis of a recently proposed phenomenological model 
characterized by radially-dependent cosmic-ray (CR) transport properties. We show that the astrophysical spectrum measured by IceCube experiment can be well explained adding to the diffuse Galactic neutrino flux (obtained with this new model) a extra-Galactic component derived from the astrophysical muonic neutrinos reconstructed in the Northern hemisphere. A good agreement between the expected astrophysical neutrino flux and the IceCube data is found for the full sky as well as for the Galactic plane region.
}
\maketitle
\section{Introduction}
\label{intro}
In the latest years the IceCube collaboration opened the era of neutrino astronomy detecting a astrophysical neutrino excess above the expected background flux. Recently, a preliminary analysis~\cite{IC_4yr_ICRC} based on 
four years experiment data-taking ($1347$ days of livetime), reported a total of $54$ HESE between $28 TeV$ and $3 PeV$. The astrophysical $\nu$ excess inferred by IceCube with HESE events collected in three
years was fitted by a power law with index $\Gamma = - 2.3 \pm 0.3$ and a best-fit per flavor flux of $E^{2}\Phi(E)=1.5\pm0.7\times10^{-8}(E/100TeV)^{-0.3}GeVcm^{-2}s^{-1}sr^{-1}$ above 60 TeV~\citep{Aartsen:2014gkd}. The four-year data favor a 
steeper spectrum: $\Gamma = - 2.58 \pm 0.25$ and a best-fit flux $E^{2}\Phi(E)=2.2\pm0.7\times10^{-8}(E/100 TeV)^{-0.58}GeVcm^{-2}s^{-1}sr^{-1}$~\cite{IC_4yr_ICRC} for the same energy range. Although a statistically significant departure from 
isotropy cannot be claimed yet,  recent analyses~\cite{Ahlers:2015moa,Neronov:2015osa} showed that the angular distribution of HESE events allows up to $50\%$ of the full-sky astrophysical flux to have a Galactic origin. 
The Galaxy can be considered a guaranteed source of neutrinos up to a fraction of PeV energies at least; a sizable flux may either come from freshly accelerated CRs undergoing hadronic scattering with gas clumps, or 
from the hadronic interactions between the Galactic CR sea and the interstellar gas. The expected diffuse neutrino flux produced by these processes is strongly dependent on the CR transport scenario. In this work we 
present the expected Galactic diffuse neutrino flux, computed with a recently introduced scenario ($KRA_\gamma$) considering a radially-dependent diffusion coefficient~\cite{2015arXiv150400227G}, for different regions of the sky. Respect to a standard 
scenario, with a fixed diffusion coefficient for the whole galaxy, this model predict a diffuse Galactic neutrino flux two times higher for the full sky region and up to four times higher for the Galactic plane region. However, to explain the spectra obtained with 
IceCube measurements for these regions it is necessary to account also for a extra-Galactic (EG) component. Here we assume this EG component to be isotropic and use the astrophysical muonic neutrino IceCube measurements from the Northern 
hemisphere~\cite{Aartsen:2015rwa} to obtain an estimation of this flux. In particular we obtain the EG neutrino spectral features considering the best-fit analysis performed by the IceCube collaboration~\cite{Aartsen:2015rwa} without a particular 
assumptions for the EG neutrino emitters. We show that the measured full sky neutrino spectrum as well as the derived neutrino spectrum of the Galactic plane region ($\theta\footnote{With $\theta$ we refer to the angular width of the selected region respect to the galactic plane line, in this case $\theta$ correspond to the galactic declination $b$}<7.5^{\circ}$) are well described by the sum of Galactic ($KRA_\gamma$) and EG (best-fit of $\nu_{\mu}$ from Northern hemisphere) selected $\nu$ spectra. 

\section{Diffuse Galactic neutrino spectra obtained with $KRA_\gamma$ model}
As explained in details in \cite{2015arXiv150400227G}, the $KRA_\gamma$ model, used in this work to obtain the Galactic diffuse neutrino emission, adopt an index $\delta$\footnote{$D(\rho) \propto \left(\rho/\rho_0\right)^\delta$ with $\rho$ representing 
the rigidity, and $\delta$ variable with R as $\delta(R)=a\times R +b$ where $a = 0.035~Kpc^{-1}$ and $b = 0.21$.} for the rigidity dependence of the diffusion coefficient that increases with the Galactocentric radius $R$ (implying spatially variable CR 
transport as originally suggested e.g. in~\cite{Erlykin:2012dp}), and this turns into a hardening of CR propagated spectrum and $\nu$ and $\gamma$-ray emissivity above the TeV energies in the inner Galaxy region. The model is tuned to reproduce over 
the entire sky the diffuse $\gamma$-ray emitting spectrum of the Galaxy as measured by Fermi-LAT~\cite{FermiLAT:2012aa}  as well as the local CR observables~\cite{Adriani:2011cu,Aguilar:2015ooa}. The hadronic emission computed by considering the 
KRA$_\gamma$ setup is higher than the conventional model predictions, with important implications for Galactic diffuse neutrino expectations. We first compute the $\nu_e$ and $\nu_\mu$ production spectra. For both flavors we use the 
emissivities provided in~\cite{kamae} (well tuned on accelerator and CR data) for projectile energies below $\sim 500$ TeV, while we adopt the ones provided in~\cite{Kelner:2006tc} above that energy range. Then we account for neutrino oscillations. 
Their effect is to almost equally redistribute the composition among the three flavors~\cite{Cavasinni:2006nx}.  We only consider proton and helium CRs/gas -- as for $\gamma$-rays -- since heavier nuclear species give a negligible contribution in the 
energy range we cover in this work~\cite{Kachelriess:2014oma}. Two possible cut-offs have been selected for the interacting CR: $5 PeV$ and $50 PeV$; these choices match CREAM p and He data and roughly KASCADE~\cite{KASCADE2005} 
and KASCADE-Grande data~\cite{Apel:2013dga}.

\section{Coupling a Galactic ($KRA_{\gamma}$ $\nu$) and a extra-Galactic ($\nu$ flux of the Northern hemisphere) component to explain IceCube data.}
\begin{figure}[ht]
\begin{tabular}{cc}
\includegraphics[width=0.44\textwidth]{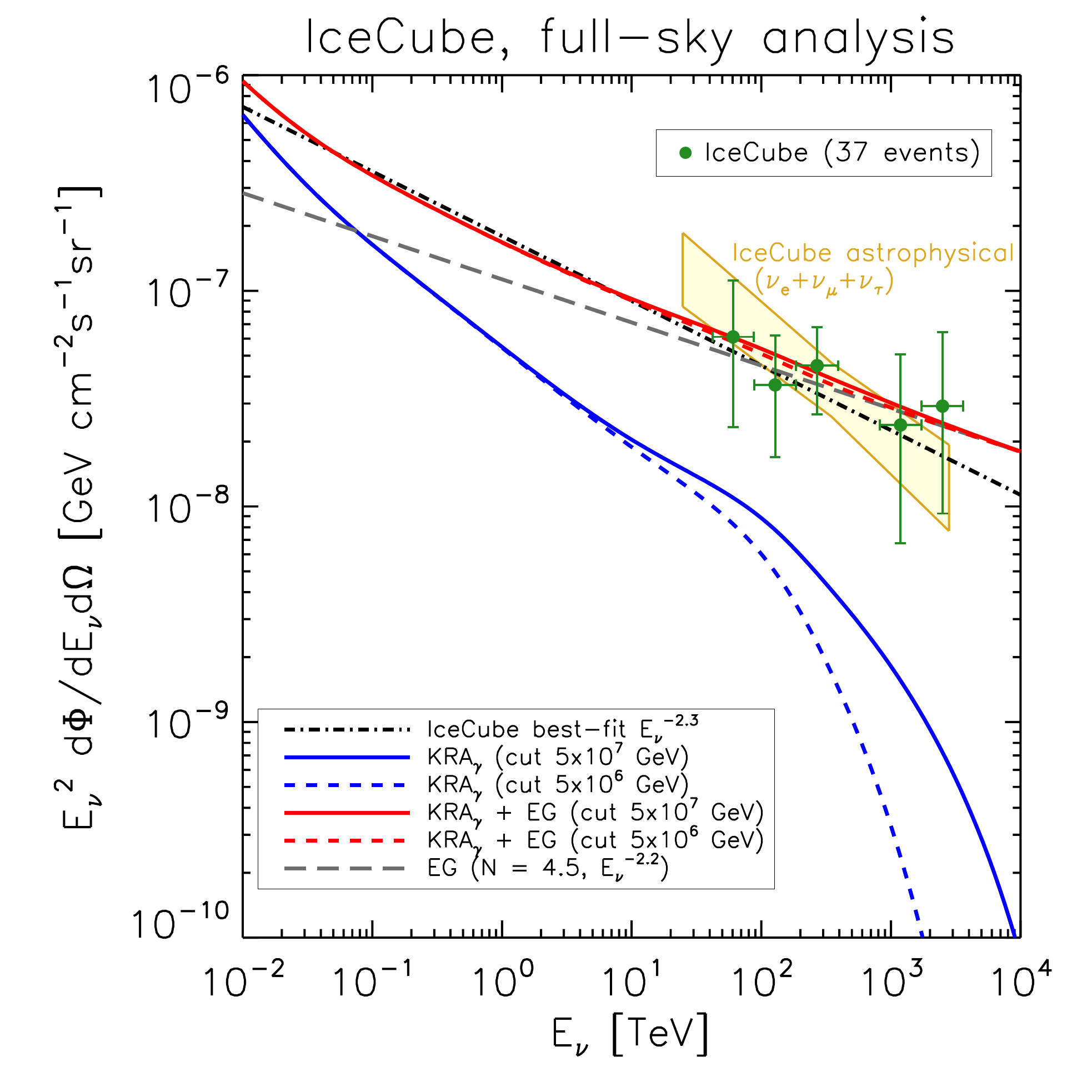}  &
\includegraphics[width=0.54\textwidth]{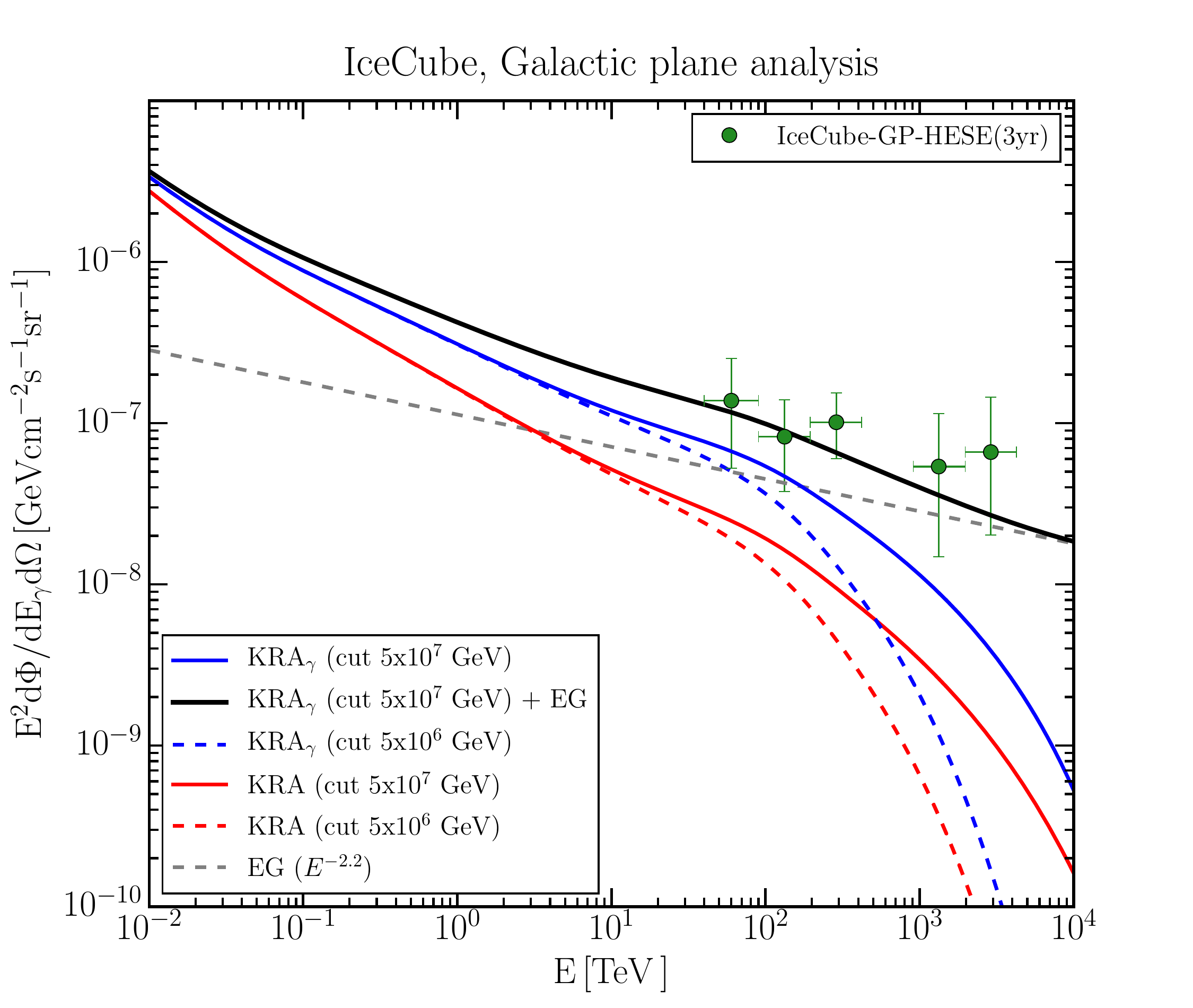}  
\end{tabular}
\caption{Comparison between IceCube spectral measurements and neutrino spectra obtained with the sum of a diffuse Galactic component (computed with the implementation of $KRA_\gamma$ model inside the  {\tt DRAGON} framework~\cite{Dragonweb}) with a isotropic EG $\nu$ spectrum derived from the $\nu_{\mu}$ measurements of the Northern hemisphere. On the left plot the full sky analysis: with blue solid and dashed lines we report the $KRA_\gamma$ $\nu$ spectra considering CR cut-offs respectively at 50 and 5 PeV. With the grey dashed line the isotropic EG component obtained from the Northern $\nu_{\mu}$ measurements. The yellow region and the green points are referred to spectral analyses obtained with 3 years of HESE events. The black dash-dotted line represents the best-fit analysis of 3 years of HESE events while the red solid line represents the sum of $KRA_\gamma$ with EG component obtained from $\nu_\mu$. On the right plot the Galactic plane analysis: with the blue and red solid and dashed thin lines we show the $KRA_\gamma$ (new scenario for CR transport) and the $KRA$ (conventional scenario for CR transport) $\nu$ spectra with the CR cut-offs at 50 and 5 PeV. The grey dashed line is the EG spectral component, while the blue solid thick line is the sum of $KRA_\gamma$ neutrinos and EG spectrum. With the green points the derived spectral points for the Galactic plane region $\theta<7.5^{\circ}$.}\label{KRA-gamma-spectra}
\end{figure}

In this work we present the interpretation of  the astrophysical neutrino spectrum measured by IceCube adding the Galactic diffuse emission, obtained with $KRA_{\gamma}$ scenario, to a diffuse EG neutrino component derived from the best-fit 
analysis of muonic neutrinos coming from the Northern hemisphere~\cite{Aartsen:2015rwa}. Since the inner Galactic plane is located in the South hemisphere, we assume the muonic neutrino flux from the Northern hemisphere ($\Phi^{\rm North}_{\nu_
\mu} = 1.7^{+ 0.6}_{- 0.8} \times 10^{-18}~\left(E/100~{\rm TeV}\right)^{-2.2 \pm 0.2}~(GeV~cm^{2}~sr~s)^{-1}$) to be a good approximation of a isotropic EG neutrino component. The IceCube best-fit spectrum~\cite{Aartsen:2015rwa} was 
multiplied by three for taking into account the sum of $\nu_{\mu}$, $\nu_{e}$ and $\nu_{\tau}$. This analysis was done for the full-sky region as well as for the whole Galactic plane inside the latitude of $\theta<7.5^{\circ}$. For the full sky 
region we compare the sum of computed neutrino fluxes ($KRA_{\gamma}$ $+$ North $\nu_{\mu}\times3$) with the IceCube spectral analyses corresponding to HESE events collected in three years~\cite{Aartsen:2014gkd, Aartsen:2015knd}. Instead for 
the Galactic plane we derived the spectral points following~\cite{Spurio:2014una} guideline. In particular we renormalized the IceCube spectral measurements~\cite{Aartsen:2014gkd} considering the neutrino events reconstructed inside the Galactic plane 
window selected ($\theta<7.5^{\circ}$). As we can see in fig. \ref{KRA-gamma-spectra} adding to the $KRA_{\gamma}$ the EG neutrino component (obtained from the $\nu_{\mu}$ of Northern hemisphere) we are able to explain the IceCube spectral 
measurements. For the full sky analysis, where the Galactic contribution is expected to be subdominant, the computed $\nu$ spectrum with $KRA_{\gamma}$ allows to reproduce the best-fit analysis of HESE events at low energies. This is 
due to the steeper and higher spectrum of $KRA_\gamma$ $\nu$ at $E<100 TeV$ respect to the considered EG spectral component. Considering the Galactic plane region the $\nu$ spectrum of $KRA_\gamma$ is even more important for explaining the 
derived IceCube spectrum. The right plot in fig. \ref{KRA-gamma-spectra} shows that taking into account a standard CR transport scenario (with a fixed diffusion coefficient for the whole galaxy) for producing diffuse Galactic $\nu$, it would be needed a 
excessive high EG spectral $\nu$ counterpart to reproduce the IceCube spectrum.

\section{Conclusions}
In the presented analysis we interpreted the astrophysical neutrino flux measured by IceCube as the sum of a computed diffuse Galactic $\nu$ component with an isotropic EG $\nu$ flux obtained directly from data. The diffuse Galactic $\nu$'s, produced 
by cosmic-ray interaction inside our Galaxy, are obtained following a new CR transport scenario, called $KRA_\gamma$, where the diffusion coefficient is assumed to be radially-dependent. Whereas the EG $\nu$ flux is obtained from a particular IceCube 
analysis. Since the inner Galactic plane is positioned in the South hemisphere, we took the $\nu_{\mu}$ flux reconstructed in the Northern hemisphere as a good estimate of the isotropic EG neutrino component. The spectral features of the EG spectrum 
correspond to the best fit parameters given by the IceCube collaboration. The sum of these two astrophysical components well reproduce the IceCube spectral measurements for the full sky region as well as for the whole Galactic plane with $
\theta<7.5^{\circ}$. For both regions we highlight the importance of using the $KRA_{\gamma}$ model instead of a standard CR transport scenario to obtain the expected Galactic diffuse neutrino flux. In a near future the IceCube-ANTARES 
common analysis of the Galactic plane can confirm the features of neutrino spectrum predicted by the $KRA_{\gamma}$ scenario and constrain the spectral characteristics of EG neutrino flux.

%\begin{figure}[ht]
%\centering
%\includegraphics[scale=0.5]{KRAgamma_hardening_ICECUBE_test_Extra_2,2.pdf}
%\caption{ciao1}
%\label{KRA-gamma-full-sky}
%\end{figure}
%\begin{figure}[ht]
%\centering
%\includegraphics[scale=0.45]{neutrino_plot__plane.pdf}
%\caption{ciao2}
%\label{KRA-gamma--plane}
%\end{figure}
%For bibliography use \cite{Aartsen:2015trq}
%\subsection{Subsection title}
%\label{sec-2}

%\begin{figure}[ht]
%\centering
%\includegraphics[width=0.8\textwidth]{KRA-gamma-IceCube-spectra.pdf}
%\caption{ciao}\label{KRA-gamma-IceCube-spec}
%\end{figure}
%
% BibTeX or Biber users please use (the style is already called in the class, ensure that the "woc.bst" style is in your local directory)
% \bibliography{bibneutrino}
%
%%%%%%%%% Non-BibTeX users please use
%%%%%%%%%
%%%%%%%%\begin{thebibliography}{}
%%%%%%%%%
%%%%%%%%% and use \bibitem to create references.
%%%%%%%%%
%%%%%%%%\bibitem{RefJ}
%%%%%%%%% Format for Journal Reference
%%%%%%%%Journal Author, Journal \textbf{Volume}, page numbers (year)
%%%%%%%%% Format for books
%%%%%%%%\bibitem{RefB}
%%%%%%%%Book Author, \textit{Book title} (Publisher, place, year) page numbers
%%%%%%%%% etc
%%%%%%%%\end{thebibliography}

\end{document}